\documentclass[12pt]{article}
\usepackage[totalheight = 23cm, totalwidth = 17cm]{geometry}

\newcommand{\be}{\begin{equation}}
\newcommand{\ee}{\end{equation}}
\newcommand{\bea}{\begin{eqnarray}}
\newcommand{\eea}{\end{eqnarray}}

\newcommand{\hf}{\frac{1}{2}}
\newcommand{\nn}{\nonumber\\}

\usepackage{graphicx}
\usepackage{epsf}
\usepackage{epsfig}
\begin{document}

\pagestyle{empty}
\begin{flushright}
{CERN-PH-TH/2006-240}\\
hep-th/0611228\\
\end{flushright}
\vspace*{5mm}
\begin{center}
{\large {\bf Non-Perturbative Formulation of Non-Critical String Models}} \\
\vspace*{1cm}
{\bf Jean Alexandre$^1$}, {\bf John~Ellis$^{1,2}$} and {\bf Nikolaos E. Mavromatos$^1$} \\
\vspace{0.3cm}
$^1$ Department of Physics, King's College, London, England \\
$^2$ Theory Division, Physics Department, CERN, CH-1211 Geneva 23, Switzerland \\

\vspace*{2cm}
{\bf ABSTRACT} \\
\end{center}
\vspace*{5mm}
\noindent
We apply to non-critical bosonic Liouville string models, characterized by a central-charge deficit
$Q$, a new non-perturbative renormalization-group technique
based on a functional method for controlling the quantum fluctuations. We demonstrate
the existence of a renormalization-group fixed point of Liouville string theory as $Q \to 0$, in which
limit the target space-time is Minkowski and the dynamics of the
Liouville field is trivial, as it neither propagates nor interacts. This calculation supports in a
non-trival manner the identification of the zero mode of the Liouville field with the target time
variable, up to a crucial minus sign.
\vspace*{5cm}
\noindent

\begin{flushleft} CERN-PH-TH/2006-240 \\
November 2006\\
\end{flushleft}
\vfill\eject
%\pagestyle{empty}
%\clearpage\mbox{}\clearpage

\setcounter{page}{1}
\pagestyle{plain}

\pagebreak

\section{Introduction}

In non-critical strings of the type discussed in~\cite{aben},
one has a conformal field theory formulated in general in a
non-critical number of spatial dimensions with a central charge deficit $Q$.
This might either assume
discrete values, as in the minimal models discussed in~\cite{aben}, or
possibly vary continuously, as in models motivated by brane
world collisions~\cite{brany}. In the latter case, the central charge
of the corresponding world-sheet $\sigma$ model describing string excitations
on the brane and in the bulk space is proportional to some power of the
relative velocity of the moving models (assuming that the collisions are adiabatic, so that
perturbative string theory applies).

In general, non-equilibrium situations in string cosmology, such as those that
may well have characterized the early Universe, can be described~\cite{brany}
at large times long after the initial cosmic catastrophe that resulted in
the departure from equilibrium, within the framework of
Liouville strings~\cite{ddk}. The latter are strings described by world-sheet
$\sigma$-models propagating in non-conformal backgrounds of, say, graviton and dilaton fields,
that are dressed by an extra world-sheet field, the Liouville mode $\phi$,
in such a way that conformal invariance is restored. This construction enables
strings to propagate in a non-critical number of space-time dimensions.

It was argued in~\cite{brany} that in some \emph{supercitical}
models, i.e., world-sheet $\sigma$models with a
central charge \emph{surplus}: $-Q^2 \equiv (C-c^*)/3 > 0$
where $c^*$ is the critical central charge of the conformal theory.
the extra Liouville dimension, i.e., the zero mode of the world-sheetLiouville field $\phi$,
can be identified with the \emph{target time}. This identification
follows from dynamical arguments on the minimization of the
effective potential of the target-space-time effective
field theory, and is exemplified by, e.g., black-hole configurations.

Generic analyses~\cite{diamand}  of cosmological models within this general framework
of Liouville cosmologies, which have been termed $Q$-cosmologies, reveals that the
asymptotic theory at large times corresponds to the conformal model of~\cite{aben}, with
a central charge given by the asymptotic constant value $Q_0$ of the central-charge deficit.
It should be noted that, in general, the central charge
of the Liouville cosmology is not a constant, but a time-dependent function,
$Q(t)$, whose form is found by solving the appropriate generalized conformal-invariance
conditions that describe the restoration of conformal invariance by the Liouville mode.

The cosmology of~\cite{aben} corresponds in target space to a linearly-expanding Universe.
However, the question arises how the geometry of the Universe evolves with time and,
in particular, whether and how this Universe exits from this expanding phase and reaches
an Minkowski space-time. The latter is the only realistic candidate for
a \emph{equilibrium} situation which may be reached
asymptotically in target time.

It was attempted in~\cite{hall} to visualize this evolving string Universe as a
world-sheet quantum Hall system, with the cosmologies of~\cite{aben}, that correspond to
various \emph{discrete} values of the central-charge deficit $Q$, being the
analogues of the conductivity plateaux of the Hall system. Transitions
between them, from one value $Q_1$ to another value $Q_2$ in, say, the discrete
series found in minimal models, would correspond to a non-conformal theory dressed
by the world-sheet Liouville mode.
According to~\cite{aben} therefore, the Universe would undergo a series of phase transitions
before reaching asymptotically the equilibrium Minkowski space-time that
corresponds to the $Q=0$ critical theory.
The question that then arises is how to describe such phase transitions
non-perturbatively on the world-sheet.

In ordinary field theory, the approach to a phase transition
is described by means of a renormalization-group flow.
An alterative to the conventional Wilsonian flow method was presented in~\cite{polonyi},
in which  a mass parameter is relaxed from some high value, where the quantum corrections
are well controlled, down to small values. This procedure was applied initially
to $\phi^4$ field theory and then to QED and some $2+1$-dimensional models.
More recently, we applied this approach to string theory, imposing a fixed ultraviolet cutoff
$\Lambda$ on the world sheet, and using the Regge slope $\alpha^{'}$
as the control parameter~\cite{alexandre}.
In this way, we found a novel fixed point of the world-sheet
$\sigma$-model describing the bosonic string in cosmological graviton and dilaton
backgrounds, which is non-perturbative in $\alpha^{'}$
and describes a novel time-dependent string cosmology.
This novel fixed point is an infrared fixed point of the Wilsonian renormalization group,
and a marginal configuration of the alternative flow.
These theories remain conformal,
and one of the non-trivial tasks in~\cite{alexandre} was to argue that the new fixed
point respects world-sheet conformal invariance.

In this paper we
extend these results to Liouville theory, using as the control parameter of the
novel renormalization flow the central-charge deficit $Q$. It is known from the
original work on linearly-expanding cosmologies in~\cite{aben} that the
central charge induces mass shifts $\propto Q$ in the spectrum of
target-space excitations: there are tachyonic mass shifts, $\Delta m^2 = -|Q^2| < 0$ for
bosons when $-Q^2>0$~\footnote{We use units where $\alpha' = 1$. We note that fermion masses
do not acquire a $Q^2$ correction,
as discussed in \cite{aben}.}. In the case of initially massless states, this tachyonic shift
would imply tachyonic excitations in the spectrum, and hence instabilities.
On the other hand, its role in generating a mass gap makes $Q$ a suitable
candidate for controlling the quantum corrections. By treating it as variable,
we can discuss transitions among various linearly-expanding cosmologies,
and eventually the transition to Minkowski space as a fixed point of the
novel renormalization flow.

\section{Non-Perturbative Flows with Respect to the Central-Charge Deficit}

The bare action for the two-dimensional world-sheet $\sigma$ model for the
bosonic string is
\be\label{liouvsmodel}
S=\int d^2\xi\left\{\frac{Q^2}{2}\partial_a\phi\partial^a\phi+\beta_Q R^{(2)}\phi+\mu^2P_B(\phi)e^\phi\right\},
\ee
where $\beta_Q$ is a function of $Q^2=\frac{c^*-C}{3}$, $c^*=25$,
and $P_B(\phi)$ is a $Q^2$-independent
bare polynomial in the Liouville field $\phi$.
The effective action $\Gamma$, which is the generating functional for the
proper graphs, is defined in the Appendix. It describes the
corresponding quantum theory, and is labelled by the parameter $Q^2$.

The target-space  Liouville field is {\it space-like} in when the
corresponding conformal theory is {\it subcritical}, i.e., is characterised by a central charge
{\it deficit}~\cite{ddk}, i.e., $Q^2 > 0$. On the other hand,
the target-space Liouville field is {\it time-like} in when the corresponding world-sheet theory is
{\it supercritical}, i.e., there is a central-charge {\it surplus}~\cite{aben}: $Q^2 < 0$.
It is the latter case that has been
employed previously~\cite{emn,diamand} to describe (non-equilibrium) string cosmologies,
which relax to equilibrium (critical-string) configurations asymptotically in target time, 
the latter being identified with the zer mode of the
time-like Liouville field. In these cosmologies the initial central charge surplus may
be provided by some catastrophic cosmic event, e.g., the collision of brane worlds in the modern
version of string theory~\cite{diamand}.

From a world-sheet field-theory point of view, the subcritical string with
central charge $C < 1$ constitutes a
well-behaved theory, where functional computations can be performed, and
the critical (scaling) exponents of the theory are {\it real}~\cite{ddk}.
For the range $1 < C < 25$ of central charges there are {\it complex} scaling exponents, and the Liouville 
theory is at {\it strong coupling}, which is not
well understood at present.
On the other hand, the supercritical Liouville theory $C > 25$, is characterised by a ghost-like field
$\phi$, since the kinetic term of the Liouville mode comes with the `wrong' (negative in our conventions) sign  (c.f. (\ref{liouvsmodel})). However, in ths theory the critical exponents are also real, and in fact this regime can be thought of as the
analytical continuation of the region where $C < 1$, with the replacement $Q \to iQ$, with the Liouville scaling exponents $\alpha$ also undergoing a similar Wick rotation: $\alpha \to i\alpha$.

In this paper we shall present a novel way of quantising the Liouville theory, adapting a method developed previously in \cite{polonyi} for ordinary field theories. There, one identifies a
parameter (control parameter) in the theory, whose changes are governed by certain flow equations, which may be constructed by following standard (non-perturbative) functional methods. The resulting
flow describes the quantum-corrected behaviour of the theory in a non-perturbative way.

The main idea of this paper is to use the central charge deficit $Q^2$ of the
Liouville theory as an appropriate control parameter. We formulate the
flow equations first in the subcritical
case, which is well defined as a field theory, and then we continue analytically to the supercritical string case with $Q^2 < 0$.

We start our analysis at $Q^2>>1$, where
the theory is classical, since the bare Lagrangian is dominated by the kinetic term and therefore describes a free theory.
The decrease of $Q^2$ then induces the appearance of quantum fluctuations,
leading to the dressed theory. It is shown in the Appendix that it is possible to derive
an exact evolution equation for $\Gamma$ with $Q^2$, which is
\be\label{evolG}
\dot\Gamma=\int d^2\xi\left\{\hf\partial_a\phi\partial^a\phi+\dot\beta_Q R^{(2)}\phi\right\}
+\hf\mbox{Tr}\left\{\frac{\partial}{\partial\xi_a}\frac{\partial}{\partial\zeta^a}
\left(\frac{\delta^2\Gamma}{\delta\phi_\xi\delta\phi_\zeta}\right)^{-1}\right\},
\ee
where a dot denotes a derivative with respect to $Q^2$. In eq.(\ref{evolG}),
quantum fluctuations are contained in the trace on the right-hand side.
This trace needs a regulator, for which we use a fixed world-sheet cutoff $\Lambda$.

Any similarity of our evolution equation (\ref{evolG}) to the exact Wilsonian renormalization
equation is only apparent, since here we consider a {\it fixed} cutoff, and look at the
flows in $Q^2$. We emphasize that eq.(\ref{evolG}) is exact and corresponds to the
resummation of all loops, even though superficially it has the structure of a one-loop
correction. The reason for this is the fact that the trace
contains the dressed parameters, and not the bare ones: thus eq.(\ref{evolG}) is a self-consistent
partial differential equation for $\Gamma$, which describes the full quantum theory.

In order to obtain physical information from the evolution equation (\ref{evolG}), one has to assume a functional dependence of the effective action $\Gamma$.
Therefore, we consider the following Ansatz:
\be\label{ansatz}
\Gamma=\int d^2\xi\left\{ \frac{Z_Q}{2}\partial_a\phi\partial^a\phi+\beta_Q R^{(2)}\phi +
\mu^2P_Q(\phi)e^\phi\right\},
\ee
where $Z$ is a $Q^2$-dependent wave-function renormalization, and $P_Q(\phi)$ is a
$Q^2$-dependent function of $\phi$. The form (\ref{ansatz}) is dictated by conformal invariance.
Note that we do not expect quantum corrections for $\beta_Q$, since no term linear in $\phi$ is generated by the trace in eq.(\ref{evolG}).
It is shown in the Appendix that the Ansatz (\ref{ansatz}), inserted into eq.(\ref{evolG}),
leads to the following evolution equations:
\bea\label{evoleqs}
\dot Z_Q&=&1\\
\dot P_Q(\phi)&=&-\frac{P_Q(\phi)+2P_Q^{'}(\phi)+P_Q^{''}(\phi)}{8\pi Z_Q^2}
\ln\left(1+\frac{Z_Qe^{-\phi}\Lambda^2/\mu^2}{P_Q(\phi)+2P_Q^{'}(\phi)+P_Q^{''}(\phi)}\right),
\nonumber
\eea
where a prime denotes a derivative with respect to $\phi$.

We observe that $Z$ remains classical and does not receive any quantum corrections.
Since the constant of integration in the evolution equation for $Z(Q)$
is absorbed into the critical value of the central charge, we find simply that $Z_Q=Q^2$
and the resulting evolution equation for $P$ is
\be\label{evolP}
\dot P_Q(\phi)=-\frac{P_Q(\phi)+2P_Q^{'}(\phi)+P_Q^{''}(\phi)}{8\pi Q^4}
\ln\left(1+\frac{Q^2e^{-\phi}\Lambda^2/\mu^2}{P_Q(\phi)+2P_Q^{'}(\phi)+P_Q^{''}(\phi)}\right).
\ee
We observe that there is  only one exactly-marginal configuration, namely one with
$\dot P=0$, which must have $P+2P^{'}+P^{''}=0$. The solution for $P$ is then
\be
P(\phi)=(C_1+C_2\phi)e^{-\phi},
\ee
where $C_1,C_2$ are $Q^2$-independent constants. This solution corresponds to a linear potential
\be\label{potfp}
\mu^2P(\phi)e^\phi=\mu^2(C_1+C_2\phi),
\ee
which could have been expected, since this form does not generate quantum fluctuations,
and therefore should not depend on $Q^2$.

\section{Solution in the Case $P_B(\phi)=1$}

In the case where the bare potential term is $\mu^2 e^\phi$, it is known that the
effective potential is of the form $\mu^2_R\exp(g_R\phi)$, where $\mu^2_R$ and $g_R$
are renormalized parameters~\cite{Jackiw}. We indeed find a solution of (\ref{evolP})
if we consider the following Ansatz for the effective Liouville-mode  potential $V(\phi)$:
\be\label{ansatzP}
 V(\phi) =  \mu^2 P_Q(\phi)e^\phi, \qquad
P_Q(\phi)=\eta_Q\exp(\varepsilon_Q\phi),
\ee
where $\eta_Q$ and $\varepsilon_Q$ are functions of $Q^2$. Since the limit $Q^2\to\infty$
corresponds to the classical theory, the corresponding limits for these functions are
$\eta_\infty=1$ and $\varepsilon_\infty=0$, which we use
as initial conditions when we integrate their evolution equations.
In order to check that the Ansatz (\ref{ansatzP}) is indeed correct, we consider
separately the two cases of large $Q^2 \gg 1$ and $Q^2\to 0$.

\subsection{Large $Q^2$}

If we insert the Ansatz (\ref{ansatzP}) into eq.(\ref{evolP}), we obtain
\be\label{expLambda}
\dot\eta_Q+\eta_Q\dot\varepsilon_Q\phi=\frac{\eta_Q(1+\varepsilon_Q)^2}{8\pi Q^4}\left\{
-\ln\left(\frac{Q^2\Lambda^2}{\mu^2\eta_Q(1+\varepsilon_Q)^2}\right)+(1+\varepsilon_Q)\phi
+{\cal O}\left(\frac{\mu^2}{Q^2\Lambda^2}\right)\right\},
\ee
where we need the condition $Q^2\Lambda^2>>\mu^2$ for the Ansatz (\ref{ansatzP}) to be
consistent. Indeed, after the expansion in $\mu^2/(Q^2\Lambda^2)$, one is left with a constant
and a term linear in $\phi$, which can then be identified with the left-hand side of
eq.(\ref{expLambda}), leading to
\bea\label{epsilonqeq}
\dot\varepsilon_Q&=&\frac{(1+\varepsilon_Q)^3}{8\pi Q^4}\nn
\dot\eta_Q&=&\frac{\eta_Q(1+\varepsilon_Q)^2}{8\pi Q^4}
\left\{-\ln\left(\frac{Q^2\Lambda^2}{\mu^2(1+\varepsilon_Q)^2}\right)
+\ln(\eta_Q)\right\}.
\eea
The evolution equation for $\varepsilon_Q$ can easily be solved to yield
\be\label{solepsilon}
1+\varepsilon_Q=\sqrt{\frac{4\pi Q^2}{4\pi Q^2+1}},
\ee
and we stress again that this solution is not the result of a loop expansion, but is exact in the
framework of the Ansatz (\ref{ansatz}).

The solution (\ref{solepsilon}) leads to the following equation for $\eta_Q$:
\be\label{evoleta}
\frac{\dot\eta_Q}{\eta_Q}=\frac{-1}{2Q^2(4\pi Q^2+1)}\left\{\ln\left(\frac{\Lambda^2}{\mu^2}\right)
+\ln\left(Q^2+\frac{1}{4\pi}\right)-\ln(\eta_Q)\right\},
\ee
for which one can find an approximate solution if $Q^2>>1$, where $\eta_Q\simeq 1$.
We have then, for a {\it fixed} cutoff $\Lambda$, and keeping the dominant contributions,
\be
\dot\eta_Q\simeq -\frac{\ln(Q^2)}{8\pi Q^4},
\ee
which leads to the following dominant behaviour
\be
\eta_Q\simeq 1+\frac{\ln(Q^2)}{8\pi Q^2}.
\ee
In the limit where $Q^2>>1$ and from eq.(\ref{solepsilon}), we also have
$\varepsilon_Q\simeq -1/(8\pi Q^2)$, so that
the effective potential is finally
\be\label{pqfinal}
V(\phi) = \mu^2P_Q(\phi)e^\phi\simeq\mu^2\left(1+\frac{\ln(Q^2)}{8\pi Q^2}\right)
\exp\left\{\left(1-\frac{1}{8\pi Q^2}\right)\phi\right\}~.
\ee

\subsection{Limit $Q^2\to 0$}

In the limit where $Q^2\to 0$, the expansion (\ref{expLambda}) is not valid any more, and one
has to start from the original equation (\ref{evolP}). An expansion in $Q^2$ for a {\it fixed}
cutoff $\Lambda$ then gives
\be\label{expQto0}
\dot\eta_Q+\eta_Q\dot\varepsilon_Q\phi=-\frac{\Lambda^2/\mu^2}{8\pi Q^2}\exp\left\{-(1+
\varepsilon_Q)\phi\right\} +{\cal O}(1).
\ee
For the ansatz (\ref{ansatzP}) to be consistent, we consider an expansion in $\phi$ of the previous
equation, and identify the powers of $\phi$ to obtain
\bea\label{epsilonqeq0}
\dot\varepsilon_Q&=&\frac{\Lambda^2/\mu^2}{8\pi Q^2}\frac{1+\varepsilon_Q}{\eta_Q}\nn
\dot\eta_Q&=&-\frac{\Lambda^2/\mu^2}{8\pi Q^2}.
\eea
These equations can easily be integrated to give
\bea\label{limitpotqzero2}
1+\varepsilon_Q&\simeq &\left|\ln(Q^2)\right|^{-1}\nn
\eta_Q&\simeq &\frac{\Lambda^2/\mu^2}{8\pi}\left|\ln(Q^2)\right|,
\eea
where we have kept only the contributions that are dominant in $Q^2$.
Note that $1+\varepsilon_Q\to 0$, which is consistent with the expansion of the exponential function
appearing in eq.(\ref{expQto0}). Finally, the effective potential behaves as
\be\label{limitpotqzero}
V(\phi) = \mu^2P_Q(\phi)e^\phi\simeq\frac{\Lambda^2}{8\pi}\left|\ln(Q^2)\right|
\exp\left\{\frac{\phi}{\left|\ln(Q^2)\right|}\right\}
\simeq\frac{\Lambda^2}{8\pi}\left|\ln(Q^2)\right|,
\ee
and therefore goes to a (divergent) constant when $Q^2\to 0$. As a result, this limit consists
of a trivial theory, where the field $\phi$ neither propagates nor interacts. In this limit the
quantum fluctuations, from which $\varepsilon_Q$ is generated, are strong enough to cancel the
classical potential. This becomes visible in the present scheme because it is
non-perturbative.

\section{Conformal Invariance}

One of the most important properties of the Liouville field $\phi$  is the restoration of the conformal invariance of world-sheet vertex operators after Liouville dressing~\cite{ddk}, such that the Liouville-dressed world-sheet
theory, incorporating the extra dynamics of the Liouville mode $\phi$,  is conformally invariant.

Before commencing our discussion, we recall that
it is customary~\cite{ddk} to renormalize the Liouville field so that it has a canonically-normalized
kinetic term:
\bea
\phi \longrightarrow  \hat\phi \equiv |Q| \phi .
\label{norm}
\eea
For a world-sheet ($\Sigma$) vertex operator $V_i$ that deforms a fixed-point theory with action $S^*$:
\bea
 S_{\rm deform} = S^*  + g^i\int_\Sigma V_i ,
\label{deformed}
\eea
the Liouville-dressing procedure~\cite{ddk} is defined by coupling the Liouville mode $\phi$, with action $S \equiv S_L$  (\ref{liouvsmodel}), to (\ref{deformed}) as follows:
\bea \label{liouvilledressing}
S_{\rm deform,Liouville} = S^* +  S_L + g_i \int_\Sigma  e^{\alpha_i\hat\phi} V_i ,
\eea
where we have used the canonically-normalized field $\hat\phi$ (\ref{norm}).

The Liouville anomalous dimension terms $\alpha_i$ are such that, if the
deformed subcritical theory has central-charge deficit $Q^2 >0 $, then the dressed deformation in
(\ref{liouvilledressing})  $e^{\alpha_i \hat\phi} V_i $
is conformally invariant, provided that,
\bea
\alpha_i \left(\alpha_i + Q \right) = -(2 - \Delta_i) ~, \qquad Q^2 > 0 ~({\rm subcritical~strings}) ,
\label{cicond}
\eea
where $\Delta_i$ is the conformal (scaling) dimension of the operator $V_i$, and thus $\Delta_i -2$ is the scaling dimension. The relative signs are appropriate for the subcritical string case $Q^2 > 0$ of interest to us in this section, and are such that the Liouville dimension $\alpha_i$ and $Q$ are {\it real}.
The presence of the $Q$ term arises because of the appearance of the central charge deficit $Q$ in the world-sheet curvature term of the perturbative Liouville action~\cite{ddk}.

In the model at hand, the only deformation we considered explicitly was that implied by the identity operator on the world-sheet, namely the two-dimensional cosmological constant, which leads, in the quantum theory, to the effective Liouville potential term (\ref{ansatzP}). This corresponds to the case
with $\Delta_i =0$ in (\ref{cicond}). Moreover, in our (non-perturbative) quantum theory, the r\^ole of
the Liouville anomalous dimension is played by $(1 + \varepsilon _Q)/Q$, where the numerator is the exponent in (\ref{ansatzP}), whilst the r\^ole of the central charge deficit $Q$ in (\ref{cicond}), namely
the coefficient of the world-sheet curvature term in the normalized Liouville mode case $\hat\phi$,
is provided by the function $\beta_Q/Q$. Thus conformal invariance should be guaranteed provided
that the following relation holds:
\bea
(1 + \varepsilon_Q) \left(1 + \varepsilon_Q + \beta_Q \right) = -2Q^2 \quad ~\Longrightarrow ~ \quad  \beta_Q = -1 - \varepsilon_Q -\frac{2Q^2}{1 + \varepsilon _Q} .
\label{betaqepsilon}
\eea
As discussed in the Appendix and in previous sections, our quantization procedure determines
$\varepsilon_Q$ as a function of $Q$, so as to satisfy the appropriate flow equations
(\ref{epsilonqeq}) (and (\ref{epsilonqeq0}) for the $Q^2 \to 0^+$ case), assuming a specific form of
the function $Z=Q^2$, which receives no quantum corrections. Moreover, as we have seen, in our approach the function $\beta_Q$ (which is also not renormalized) is left undetermined.
Following the above discussion (c.f. (\ref{betaqepsilon})), the requirement of conformal invariance provides an extra constraint that determines the function
$\beta_Q$ in terms of $\varepsilon_Q$, with $Z=Q^2$.

It is worth checking the consistency of this approach in the conformal limit $Q^2 \to 0^+$, where one expects the Liouville theory to decouple.
Indeed, in such a limit, the expression for $1 + \varepsilon_Q$ is provided by
(\ref{limitpotqzero2}). From (\ref{betaqepsilon}), then, we derive to leading order as $Q^2 \to 0^+$:
\bea\label{betaqeps}
\beta_Q \simeq -1 - \varepsilon_Q \simeq -\frac{1}{|\ln(Q^2)|} \to 0^- ~.
\eea
which is consistent with the decoupling of the Liouville mode in this limit, since each of the three terms
in the world-sheet action (\ref{liouvsmodel}) either vanishes $(Z, \beta_Q)$ or becomes an irrelevant (Liouville-independent) constant (as is the case with the two-dimensional cosmological constant term).

In a similar spirit, the limit $Q^2 \gg 1$ can also be studied analytically. To this end, we first notice that
the relation (\ref{betaqepsilon}) is generic and applies to all ranges of $Q^2$.
In the large-$Q^2$ case, $\varepsilon_Q \simeq - 1 / 8\pi Q^2$, and
\bea\label{final2}
\beta_Q \simeq -2Q^2+ {\cal O}(1) < 0, \quad  Q^2 \to +\infty .
\eea
We now remark that the central-charge term is not supposed to change sign during its
flow~\cite{aben,ddk}, i.e., a sub(super)critical theory should remain sub(super)critical until its reaches
an equilibrium point. From
(\ref{betaqeps}), (\ref{final2}) we observe that this expectation is compatible with the above
analysis, as in both limits $\beta_Q < 0$.

\section{Case with $Q^2 < 0$: Intepretation of the Liouville Mode as Target Time}
\label{sec:time}

As mentioned above, the region of central charges for which $Q^2 < 0$
can be treated by analytic continuation of  the $C < 1$ case,
where formally $Q \to iQ$ and the Liouville scaling dimensions $\alpha \to i\alpha$. In this case, the exponents of the Liouville effective potential terms (\ref{ansatzP}), where - as we have discussed in the previous section- $1 + \varepsilon_Q$ plays  the r\^ole of a Liouville scaling dimension
for the identity operator on the world-sheet, remain {\it real}.

From a target-space-time viewpoint, in this r{\' e}gime the
Liouville-mode is time-like, and thus
its world-sheet zero mode can be interpreted as the target time~\cite{emn,diamand}.
In this case, the effective potential term in the Liouville action corresponds in general to a
cosmological tachyonic-field instability. However, as we have seen in (\ref{limitpotqzero}),
in the limit $|Q^2| \to 0^+$ the effective potential term becomes a constant independent of the
Liouville field, so the instability disappears and the target-space theory is stabilized.
The remaining part of the Section addresses some subtleties in these arguments, that arise
because the target time is actually identified~\cite{emn} (up to a sign) with a
renormalized Liouville mode $\equiv |Q| \phi$, and this renormalization is singular in the limit $|Q^2| \to 0^+$.

As already mentioned, it is customary~\cite{ddk} to renormalize the Liouville field so that it has a canonically normalized
kinetic term.
It is in the normalized form $\hat\phi$ (\ref{norm}) that the properties of the Liouville mode as a field
restoring conformal symmetry in non-critical world-sheet $\sigma$-model theories are best
studied~\cite{ddk,aben}.

If we had used this normalization from the beginning, the only term in the two-dimensional
effective action depending explicitly on the control parameter $Q$ would have been that coupled
to the world-sheet curvature, which depends linearly on the normalized Liouville field, and thus
does not generate any quantum corrections. However, having derived the effective potential
(\ref{limitpotqzero}) above, we can now insert the correctly  normalized Liouville mode and then
take the limit $|Q^2| \to 0^+$. In this case, the quantum-corrected potential,
expressed in terms of the normalized field $\hat\phi$, becomes:
\be\label{limitpotqzeronorm}
\mu^2P_Q(\hat\phi)e^{\hat\phi}\simeq\frac{\Lambda^2}{8\pi}\left|\ln|Q^2|\right|
\exp\left\{\frac{\hat\phi}{\left|Q \ln|Q^2|\right|}\right\} .
\ee
Notice first that, upon the above-mentioned complex 
continuations $Q \to iQ$ and $(1 + \varepsilon_Q) \to i(1 + \varepsilon_Q)$ in order to discuss formally the supercritical $Q^2 < 0$ case, the exponent of the effective potential remains real.
We then see that the limit $|Q^2| \to 0^+$ leads to divergent terms in the branch
$\hat\phi \in \left(0,~+\infty)\right)$, while such terms become zero in the branch
$\hat\phi \in \left(-\infty,~0)\right)$.

As already mentioned, the quantity that is actually identified~\cite{emn} as
the target time $t$ in supercritical string theories with $Q^2 < 0$ is {\it minus}
the world-sheet zero mode, $\hat\phi$, so that
\bea\label{minustime}
\hat\phi = - t~.
\eea
This identification can be derived  by using conformal field theory on the world sheet, as
described briefly below~\footnote{It may also be imposed dynamically
in certain concrete examples  of Liouville-time cosmologies involving colliding brane
worlds~\cite{gravanis}. In the latter case, the identification (\ref{minustime}) is enforced for
energetic reasons, specifically the minimization of the effective potential of the target-space theory.}.

This implies that, for the flow of cosmological time: $t \to + \infty$,
only the branch $\hat\phi \in (-\infty, ~ 0)$ is of physical relevance, which leads to a stable
target-space-time theory in the limit $|Q^2| \to 0^+$ of the full quantum theory, for the reasons
explained above. This target space stability, expressed via the disappearance of the tachyonic
modes and the vanishing of the tachyonic mass shifts $\Delta m^2=-|Q^2| < 0$ that characterize
the bosonic string states in~\cite{aben}, constitutes a physical argument in favour of the r\^ole of the
limit $|Q^2| \to 0^+$ as the final point of the flow with respect to the central charge in our approach.

For completeness, we review here briefly the derivation of the result (\ref{minustime}) from a
conformal-field-theory analysis. First of all, we note that even after quantum corrections, as our
analysis in Section 3 has shown, the effective potential
assumes the form (\ref{ansatzP}). From a world-sheet field-theory point
of view, this corresponds to a vertex operator of a Liouville-dressed cosmological constant term,
$V(z) =e^{\alpha\hat\phi}$, where $z$ is a complex world-sheet coordinate and
$\alpha (=\varepsilon _Q)$ is a constant, depending on the central-charge deficit $Q$, which plays
the r\^ole of the Liouville anomalous dimension~\cite{ddk}. More generally, one may consider
Liouville-dressed vertex operators $V_i^L \sim e^{\alpha_i\hat\phi} V_i$, where $\alpha_i$
is the corresponding Liouville anomalous dimension.
The N-point correlation functions of the world-sheet vertex operators $V_i$
can be evaluated by first performing the integration over the world-sheet Liouville zero mode. This
leads to expressions of the form:
\be
<V_{i_1} \dots V_{i_N} >_\mu = \Gamma (-s) \mu ^s
<(\int d^2z \sqrt{{\hat \gamma }}e^{\alpha\hat\phi })^s 
\tilde V_{i_1} \dots {\tilde V}_{i_N} >_{\mu =0} ,
\label{C12}
\ee
where the ${\tilde V}_i$ have the Liouville zero mode removed, $\mu$ is a
scale related to the world-sheet cosmological constant,
and $s$ is the sum of the anomalous dimensions of the
$V_i~:~s=-\sum _{i=1}^{N} \frac{\alpha _i}{\alpha } - Q/\alpha$.
As it stands, the $\Gamma (-s)$ factor implies that (\ref{C12})
is ill-defined for $s=n^+ \in Z^+$. Such cases include physically interesting
Liouville models, such as those describing matter scattering  off a two-dimensional
($s$-wave four-dimensional) string black hole~\cite{emn}, when it is excited to a `massive'
(topological) string state corresponding to a positive integer value for $s=n^+ \in {Z}^+$.
Similar divergent expressions are met in general Liouville theory when computing the
correlation functions by analytic continuation of the central charge of
the theory, so that the sums $s$ over Liouville anomalous dimensions acquire positive integer
values~\cite{goulian}. This  also leads to ill-defined $\Gamma (-s)$ factors in the appropriate
analytically-continued correlators.

\begin{figure}[t]
\begin{center}
\epsfig{figure=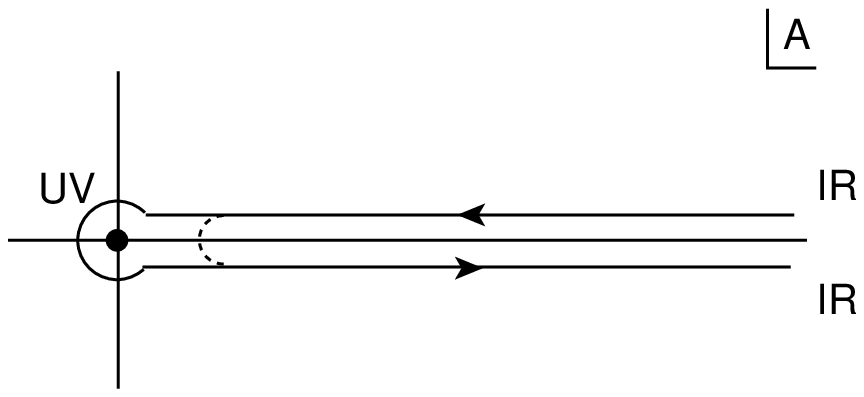,width=0.3\linewidth}
\end{center}
%\vspace{2.0in}
\caption{ {\it The solid line is the
the Saalschutz contour in the complex
area ($A$) plane, which is used to
continue analytically the prefactor
$\Gamma (-s)$ for $ s \in Z^+$. It has been used in
conventional quantum field theory to relate dimensional
regularization to the Bogoliubov-Parasiuk-Hepp-Zimmermann
renormalization method. The dashed line denotes the
regularized contour, which avoids the ultraviolet
fixed point $A \rightarrow 0$, which is used in the closed time-like path
formalism.}}
\label{fig:ctp}
\end{figure}

Constraining the world-sheet area $A$ at a fixed value~\cite{ddk}, one can use the
following integral representation for $\Gamma (-s)$:
\bea
\Gamma (-s)=\int dA e^{-A} A^{-s-1} ,
\label{integralA}
\eea
where $A$ is the covariant area of the world-sheet. In the
case $s=n^+ \in {Z}^+$ one can then regularize by analytic continuation,
replacing (\ref{integralA}) by an integral along the Saalschutz contour
shown in Fig. \ref{fig:ctp}~\cite{kogan2,emn}.
This is a well-known method of regularization
in conventional field theory, where integrals of
forms similar to (\ref{integralA}) appear in terms of
Feynman parameters.

A similar regularization was used to prove the equivalence
of the Bogolubov-Parasiuk-Hepp-Zimmerman renormalization prescription
with dimensional regularization in ordinary gauge field theories~\cite{BPZ}.
One result of such an analytic continuation is the
appearance of imaginary parts in the respective correlation functions,
which in our case are interpreted~\cite{kogan2,emn}
as renormalization-group instabilities of the system.

\begin{figure}[t]
\begin{center}
\epsfig{figure=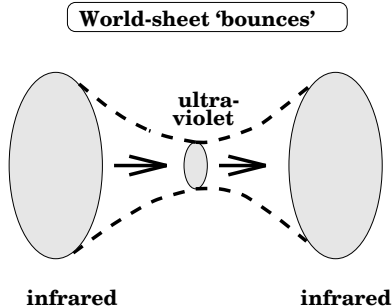,width=0.3\linewidth}
\end{center}
%\vspace{3.0in}
\caption {Schematic repesentation
of the evolution of the world-sheet area as the renormalization-group scale moves along the
contour of Fig.~\ref{fig:ctp}.}
\label{fig:wsflow}
\end{figure}

Interpreting
the latter as an actual time flow, with the identification of the
(world-sheet) zero mode with the target time~\cite{emn},
we then interpret the contour of Fig.~\ref{fig:ctp} as implying
evolution of the world-sheet area in both
(negative and positive) directions of time as seen in
Fig.~\ref{fig:wsflow}, i.e.,
infrared fixed point  $\to$ ultraviolet fixed point $\to$ infrared fixed point.
In each half of the world-sheet diagram of fig. \ref{fig:wsflow},
the Zamolodchikov $C$ theorem~\cite{zam}
tells us that we have an
irreversible Markov process.

This in turn implies a `bounce' interpretation
of the renormalization-group flow of Fig.~\ref{fig:wsflow}, in which
the infrared fixed point with large world-sheet area $A \to \infty$ is a `bounce' point,
similar to the corresponding picture in point-like field theory~\cite{coleman}.
Therefore, the physical flow of time $t$
is taken to be opposite to the conventional
renormalization-group flow, i.e., from the infrared
to the ultraviolet ($A \to 0$) fixed point on the world sheet.
In terms of the world-sheet zero mode of the Liouville field $\hat\phi_0$, we have
$\hat\phi_0 \sim {\rm ln}A \in \left(0,~-\infty \right)$. Our analysis in the previous Section
shows that the effective potential term (\ref{limitpotqzeronorm}) vanishes in the limit $Q^2 \to 0$,
so this limit corresponds to a stable target-space theory.
We stress once more that this is consistent with the disappearance (as $|Q^2| \to 0^+$)
of tachyonic instabilities in the target-space theory, as
manifested through tachyonic mass shifts $\Delta m^2 = -|Q^2| < 0$ of initially (i.e, before
Liouville dressing)  massless
target-space excitations. Thus, the analysis of this paper
reinforces the previous arguments that the (zero mode of the) world-sheet Liouville mode may
be identified (up to a sign) with the target time.

\section{Summary and Perspectives}

We have demonstrated in this paper how a novel renormalization-group technique for controlling
quantum effects by relaxing a mass parameter can be used to obtain non-perturbative results for
non-critical string models. We have studied the behaviour of Liouville string theory as a
function of the departure from criticality, as parametrized by the central-charge deficit $Q$.
We have identified a renormalization-group fixed point in the limit $Q^2 \to 0^+$, in which the
dynamics of the Liouville field becomes trivial, as it neither propagates nor interacts, and
the target space-time is of Minkowski type (in the supercritical string case).
We have shown that the resulting theory is free of tachyonic instabilities in target space in the limit
$|Q^2| \to 0^+$. This analysis supports the previous identification~\cite{emn} of the (zero mode
of the) Liouville mode with the target time.

This approach may in the future be used to discuss the transitions between linear-dilaton cosmological
models with different values of $Q$, and ultimately the transition to an asymptotic state. It has
been shown previously that $Q$ corresponds to the vacuum energy in conventional
field-theoretical models of cosmological inflation~\cite{diamand,emn}. The transition from scalar field energy to
relativistic particles has bee studied extensively within that framework, and our approach
provides a framework for addressing such cosmological phase transitions in string theory.

Another area where this technique may be applied is the Quantum Hall effect (QHE). The
different values of $Q$ correspond to different Hall conductivity plateaux, and our approach
can be used to discuss transitions between these plateaux. The analogy between string
cosmology and black-hole physics, on the one hand, and the QHE, on the other hand, has
been advertised previously~\cite{hall}. The novel renormalization-group described here provides a
tool that can be used to quantify this relationship.

\section*{Acknowledgements}

The work of J.E. and N.E.M. was supported in part by the European Union through the Marie Curie Research and Training Network UniverseNet (MRTN-CT-2006-035863).

\section*{Appendix}

We review here the construction of the effective action $\Gamma$ and derive
the equation describing its evolution with $Q$. For reasons explained in the text, we restrict ourselves
to the subcritical string case $Q^2 \propto c^* -C > 0$. The supercritical string case $Q^2 < 0$ is treated formally by means of analytic continuation.
In terms of the microscopic field $\tilde\phi$, the bare action is
\be
S=\int d^2\xi\left\{\frac{Q^2}{2}\partial_a\tilde\phi\partial^a\tilde\phi+\beta_Q R^{(2)}\tilde\phi
+\mu^2P_B(\phi)e^{\tilde\phi}\right\},
\ee
The partition function, namely the functional of the source $j$, is defined as
\be
{\cal Z}[j]=\int{\cal D}[\tilde\phi]\exp\left(-S-\int d^2\xi~j\tilde\phi\right),
\ee
and is related to the functional $W$ that generates connected graphs by
\be
W[j]=-\ln{\cal Z}[j].
\ee
The classical field $\phi$ is defined by differentiation of $W$ with respect to the source $j$,
and we have
\bea\label{diffW}
\frac{\delta W}{\delta j_\xi}&=&-\frac{1}{{\cal Z}}\frac{\partial {\cal Z}}{\partial j_\xi}
=\frac{<\tilde\phi_\xi>}{{\cal Z}}=\phi_\xi\nn
\frac{\delta^2 W}{\delta j_\xi\delta j_\zeta}&=&\phi_\xi\phi_\zeta-
\frac{<\tilde\phi_\xi\tilde\phi_\zeta>}{{\cal Z}},
\eea
where the quantum vacuum expectation value is
\be
<\cdot\cdot\cdot>=\int{\cal D}[\tilde\phi](\cdot\cdot\cdot)\exp\left(-S-\int d^2\xi~j\tilde\phi\right).
\ee
The effective action $\Gamma$, a functional of the classical field $\phi$, is introduced
as the Legendre transform of $W$:
\be
\Gamma[\phi]=W[j]-\int j\phi,
\ee
where the source $j$ has to be seen as a functional of $\phi$. The functional derivatives of $\Gamma$
are then
\bea
\frac{\delta\Gamma}{\delta\phi_\xi}&=&-j_\xi\nn
\frac{\delta^2\Gamma}{\delta\phi_\xi\delta\phi_\zeta}&=&
-\left(\frac{\delta\phi_\xi}{\delta j_\zeta}\right)^{-1}=
-\left(\frac{\delta^2 W}{\delta j_\xi\delta j_\zeta}\right)^{-1}.
\eea
From eqs.(\ref{diffW}), the equation describing the evolution of $W$ with $Q^2$ is
\bea
\dot W&=&\frac{1}{{\cal Z}}\int d^2\xi\int d^2\zeta\left\{
\hf\frac{\partial}{\partial\xi^a}\frac{\partial}{\partial\zeta_a}
\left<\tilde\phi_\xi\tilde\phi_\zeta\right>+\dot\beta_Q R^{(2)}\left<\tilde\phi_\xi\right>\right\}
\delta^{(2)}(\xi-\zeta)\nn
&=&\int d^2\xi \left\{\hf\partial_a\phi\partial^a\phi+\dot\beta_Q R^{(2)}\phi\right\}
-\hf\mbox{Tr}\left\{\frac{\partial}{\partial\xi_a}\frac{\partial}{\partial\zeta^a}
\left(\frac{\delta^2W}{\delta j_\xi\delta j_\zeta}\right)\right\}.
\eea
In order to find the evolution equation for $\Gamma$, one should remember that its independent
variables are $Q$ ad $\phi$, so that
\be
\dot\Gamma=\dot W+\int d^2\xi~\frac{\delta W}{\delta j}\partial_Qj-\int d^2\xi~\partial_Qj\phi=\dot W.
\ee
Using the previous results, finally we have
\be\label{evolGApp}
\dot\Gamma=\int d^2\xi\left\{\hf\partial_a\phi\partial^a\phi+\dot\beta_Q R^{(2)}\phi\right\}
+\hf\mbox{Tr}\left\{\frac{\partial}{\partial\xi_a}\frac{\partial}{\partial\zeta^a}
\left(\frac{\delta^2\Gamma}{\delta\phi_\xi\delta\phi_\zeta}\right)^{-1}\right\}.
\ee

\vspace{0.5cm}

In order to extract physical quantities from the evolution equation (\ref{evolGApp}),
we assume the following functional dependence of the effective action:
\be
\Gamma=\int d^2\xi\left\{\frac{Z_Q}{2}\partial_a\phi\partial^a\phi+\beta_Q R^{(2)}\phi+
\mu^2P_Q(\phi)e^\phi\right\}.
\ee
We have then
\bea
\frac{\delta^2\Gamma}{\delta\phi_\xi\delta\phi_\zeta}&=&\left\{Z_Q\partial_a\partial^a+
U^{''}_Q(\phi)\right\} \delta^{(2)}(\xi-\zeta),\\
&&\mbox{where}~~~U_Q(\phi)=\mu^2P_Q(\phi)e^\phi,\nonumber
\eea
and a prime denotes a derivative with respect to $\phi$.
For the evolution of $P$ only, it would be enough to insert in the evolution equation (\ref{evolGApp})
a constant field $\phi_0$. But in order to derive the evolution of $Z_Q$, one needs a varying field and we
consider thus $\phi=\phi_0+2\rho\cos(k\xi)$, where $k$ is some fixed momentum. If ${\cal A}$ denotes the surface area of the world sheet, the effective action then reads
\be\label{expGamma}
\Gamma={\cal A}\left(Z\rho^2 k^2+\beta_Q R^{(2)}\phi_0+U_Q(\phi_0)
+\frac{1}{2}\rho^2U^{''}_Q(\phi_0)+{\cal O}(\rho^3)\right),
\ee
so that the evolution equation for $U$ is obtained by identifying the $k$-independent terms in
eq.(\ref{evolGApp}), and the evolution equation for $Z$ by identifying the
terms proportional to $\rho^2k^2$.

The second derivative of the effective action reads for this configuration $\phi$, in Fourier components,
\bea
\frac{\delta^2\Gamma}{\delta\phi_p\delta\phi_q}
&=&\left(Z_Qp^2+U_Q^{''}(\phi_0)\right)(2\pi)^2\delta^{(2)}(p+q)\\
&&+\rho^2U^{'''}_Q(\phi_0)(2\pi)^2\left[\delta^{(2)}(p+q+k)+\delta^{(2)}(p+q-k)\right]+{\cal O}(\rho^3).\nonumber
\eea
The inverse of this matrix with components $p,q$ is computed using the following expansion
\be\label{expansion}
(A+B)^{-1}=A^{-1}-A^{-1}BA^{-1}+A^{-1}BA^{-1}BA^{-1}+\cdot\cdot\cdot,
\ee
where $A$ is a diagonal matrix with indices $p,q$, and $B$ is off-diagonal and
proportional to $\rho^2$.
In the previous expansion, the term linear in $A^{-1}$ is independent of $\rho,k$. It leads to the
evolution of $U$, and makes the following contribution to the trace which appears in
eq.(\ref{evolGApp}):
\bea\label{quadratic}
&&{\cal A}\int\frac{d^2p}{(2\pi)^2}\frac{p^2}{Z_Qp^2+U^{''}_Q(\phi_0)}\nn
&=&{\cal A}\frac{\Lambda^2}{4\pi Z_Q}-{\cal A}\frac{U^{''}_Q(\phi_0)}
{4\pi Z^2}\ln\left(1+\frac{Z_Q\Lambda^2}{U^{''}_Q(\phi_0)}\right).
\eea
We note that the quadratic divergence is field-independent, and therefore is irrelevant. Also,
the term linear in $B$, which appears in the expansion (\ref{expansion}), has a vanishing trace
since it is off-diagonal.
The term quadratic in $B$ in the expansion (\ref{expansion}) contains a contribution that is
proportional to $\rho^2$ and independent of $k$, which does not bring any new information,
since it corresponds to the evolution of $U^{''}$, as can be seen from eq.(\ref{expGamma}).
It also contains a contribution proportional to $\rho^2 k^2$, which leads to the evolution of
$Z$. The corresponding trace is
\bea
&&{\cal A}\rho^2\left[U^{'''}_Q(\phi_0)\right]^2
\int\frac{d^2p}{(2\pi)^2}\frac{4p^2}{\left(Z_Qp^2+U_Q^{''}(\phi_0)\right)^4}
\left(-Z_Qk^2+\frac{4Z_Q^2(kp)^2}{Z_Qp^2+U^{''}_Q(\phi_0)}\right)+{\cal O}(k^4)\nn
&=&{\cal A}\frac{\rho^2 k^2}{\pi Z_Q}\left[U^{'''}_Q(\phi_0)\right]^2\int_0^\infty dx
\left(\frac{-x}{\left(x+U_Q^{''}(\phi_0)\right)^4}+\frac{2x^2}{\left(x+U_Q^{''}(\phi_0)\right)^5}\right)+{\cal O}(k^4)\nn
&=&{\cal O}(k^4),
\eea
where we used the fact that, for any function $f(p^2)$,
\be
\int\frac{d^2p}{(2\pi)^2}(kp)^2f(p^2)=\frac{k^2}{8\pi}\int d(p^2) p^2f(p^2).
\ee
As a consequence, $Z$ does not receive quantum corrections.
Finally, the evolution equation for $P$ is found
from eqs.(\ref{evolGApp}), (\ref{expGamma}) and (\ref{quadratic})
where we disregard the field-independent quadratic divergence, to be
\be
\dot P_Q(\phi)=-\frac{P_Q(\phi)+2P_Q^{'}(\phi)+P_Q^{''}(\phi)}{8\pi Z_Q^2}
\ln\left(1+\frac{Z_Qe^{-\phi}\Lambda^2/\mu^2}{P_Q(\phi)+2P_Q^{'}(\phi)+P_Q^{''}(\phi)}\right).
\ee
The reader can now see easily why the supercritical string case $Q^2 < 0$ presents
certain problems that can be treated by analytic continuation.

Considering the case $Q^2 < 0$ and a Euclidean world sheet metric, we have
\be
\frac{\delta^2 S(\phi_0)}{\delta\phi(p)\delta\phi(q)}=
\left\{-|Q^2|(p_1^2+p_2^2)+\mu^2 e^{\phi_0}\right\}\delta^{(2)}(p+q).
\ee
The propagator is the inverse of this, and hence cannot be integrated because of the
pole, whose presence is linked to the supercriticality of the string. This pole is not the usual one corresponding to a mass. Indeed, if one returns to a
Minkowski world-sheet metric, one obtains:
\be\label{minkowsi}
\frac{\delta^2 S(\phi_0)}{\delta\phi(p)\delta\phi(q)}=
\left\{|Q^2|(p_0^2-p_1^2)+\mu^2 e^{\phi_0}\right\}\delta^{(2)}(p+q),
\ee
where $p_0=ip_2$.
One should perform another `Wick rotation' on $p_1$ in order to treat the problem properly.

Formally, these issues are resolved simply by treating the $Q^2 < 0$ case in our method by the
above-mentioned complex continuation of both $Q \to iQ$ and the Liouville scaling exponents:
$\alpha =\frac{(1 + \varepsilon_Q)}{Q} \to i\alpha $.


\begin{thebibliography}{99}

\bibitem{aben} I.~Antoniadis, C.~Bachas, J.~R.~Ellis and D.~V.~Nanopoulos,
  %``COSMOLOGICAL STRING THEORIES AND DISCRETE INFLATION,''
  Phys.\ Lett.\ B {\bf 211}, 393 (1988);
  %%CITATION = PHLTA,B211,393;%%
  %``AN EXPANDING UNIVERSE IN STRING THEORY,''
  Nucl.\ Phys.\ B {\bf 328}, 117 (1989);
  %%CITATION = NUPHA,B328,117;%%
%``Comments on cosmological string solutions,''
  Phys.\ Lett.\ B {\bf 257}, 278 (1991).
  %%CITATION = PHLTA,B257,278;%%




\bibitem{brany} J.~R.~Ellis, N.~E.~Mavromatos,
  D.~V.~Nanopoulos and A.~Sakharov,
  %``Brany Liouville inflation,''
  New J.\ Phys.\  {\bf 6}, 171 (2004)
  [arXiv:gr-qc/0407089];
  %%CITATION = GR-QC 0407089;%%
  J.~R.~Ellis, N.~E.~Mavromatos, D.~V.~Nanopoulos and M.~Westmuckett,
  %``Liouville cosmology at zero and finite temperatures,''
  Int.\ J.\ Mod.\ Phys.\ A {\bf 21}, 1379 (2006)
  [arXiv:gr-qc/0508105], and references therein.
  %%CITATION = GR-QC 0508105;%%

\bibitem{ddk} F.~David,
  %``CONFORMAL FIELD THEORIES COUPLED TO 2-d GRAVITY IN THE CONFORMAL GAUGE,''
  Mod.\ Phys.\ Lett.\ A {\bf 3}, 1651 (1988);
  %%CITATION = MPLAE,A3,1651;%%
  J.~Distler and H.~Kawai,
   ``Conformal Field Theory And 2-D Quantum Gravity Or Who's Afraid Of Joseph
  %Liouville?,''
  Nucl.\ Phys.\ B {\bf 321}, 509 (1989).
  %%CITATION = NUPHA,B321,509;%%


\bibitem{diamand}  J.~R.~Ellis, N.~E.~Mavromatos and D.~V.~Nanopoulos,
  %``A String scenario for inflationary cosmology,''
  Mod.\ Phys.\ Lett.\ A {\bf 10}, 1685 (1995)
  [arXiv:hep-th/9503162];
  %%CITATION = HEP-TH 9503162;%%
 G.~A.~Diamandis, B.~C.~Georgalas, N.~E.~Mavromatos, E.~Papantonopoulos and I.~Pappa,
  %``Cosmological evolution in a type-0 string theory,''
  Int.\ J.\ Mod.\ Phys.\ A {\bf 17}, 2241 (2002)
  [arXiv:hep-th/0107124].
  %%CITATION = HEP-TH 0107124;%%



\bibitem{hall} J.~R.~Ellis, N.~E.~Mavromatos and D.~V.~Nanopoulos,
  %``The String universe: High T(c) superconductor or quantum Hall conductor?,''
  Phys.\ Lett.\ B {\bf 296}, 40 (1992)
  [arXiv:hep-th/9209013].
  %%CITATION = HEP-TH 9209013;%%

\bibitem{polonyi} J.~Alexandre and J.~Polonyi,
%``Functional Callan-Symanzik equation,''
Annals Phys.\  {\bf 288}, 37 (2001)
[arXiv:hep-th/0010128].
%%CITATION = HEP-TH 0010128;%%

\bibitem{alexandre} J.~Alexandre, J.~Ellis and N.~E.~Mavromatos,
  %``Non-perturbative formulation of time-dependent string solutions,''
  arXiv:hep-th/0610072.
  %%CITATION = HEP-TH 0610072;%%

\bibitem{Jackiw} E.~D'Hoker and R.~Jackiw,
  %``Liouville Field Theory,''
  Phys.\ Rev.\ D {\bf 26}, 3517 (1982).
  %%CITATION = PHRVA,D26,3517;%%

\bibitem{emn} J.~R.~Ellis, N.~E.~Mavromatos and D.~V.~Nanopoulos,
  %``A Noncritical string approach to black holes, time and quantum dynamics,''
  arXiv:hep-th/9403133,  Published in Erice Subnuclear Series, 1993: pp 1-66;
  %``A microscopic Liouville arrow of time,''
  arXiv:hep-th/9805120.
  %%CITATION = HEP-TH 9805120;%%
  %%CITATION = HEP-TH 9403133;%%

\bibitem{gravanis} E.~Gravanis and N.~E.~Mavromatos,
  %``Vacuum energy and cosmological supersymmetry breaking in brane worlds,''
  Phys.\ Lett.\ B {\bf 547}, 117 (2002)
  [arXiv:hep-th/0205298].
  %%CITATION = HEP-TH 0205298;%%

\bibitem{goulian} M.~Goulian and M.~Li,
  %``Correlation functions in Liouville theory,''
  Phys.\ Rev.\ Lett.\  {\bf 66}, 2051 (1991).
  %%CITATION = PRLTA,66,2051;%%

\bibitem{kogan2} I. Kogan, Phys. Lett. B265, 269 (1991).

\bibitem{BPZ} T. Roy and A. Roy Chowdhuri,
Phys. Rev. D15 3768 (1977).

\bibitem{coleman} S. Coleman, Phys. Rev. D15 2929 (1977);
1248 (E) (1977).
\par C. Callan and S. Coleman, Phys. Rev. D16, 1762 (1977).

\bibitem{zam} A.B.~Zamolodchikov, JETP Lett.~43, 730 (1986); Sov.\
J. Nucl.\ Phys.\ 46, 1090 (1987).


\end{thebibliography}
\end{document}